\documentclass{osa-article}

\journal{oe}

\articletype{Research Article}

\begin{document}

\title{Quantum-limited discrimination of laser light and thermal light}

\author{Jonathan L. Habif,\authormark{1,2,*} Arunkumar Jagannathan,\authormark{1} Samuel Gartenstein,\authormark{1} Phoebe Amory,\authormark{1} and Saikat Guha\authormark{3}}

\address{\authormark{1}Laboratory for Quantum-Limited Information, USC Information Sciences Institute, 890 Winter St., Waltham, MA 02451 \\
\address{\authormark{2}Ming Hsieh Department of Electrical and Computer Engineering, University of Southern California, Los Angeles, CA 90089}
\authormark{3}College of Optical Sciences, University of Arizona, 1630 E. University Blvd. Tucson, AZ 85721}

\email{\authormark{*}jhabif@isi.edu} 



\begin{abstract}

Understanding the fundamental sensitivity limit of an optical sensor requires a full quantum mechanical description of the sensing task.  In this work, we calculate the fundamental (quantum) limit for discriminating between pure laser light and thermal noise in a photon-starved regime.  The Helstrom bound for discrimination error probability for single mode measurement is computed along with error probability bounds for direct detection, coherent homodyne detection and the Kennedy receiver.  A generalized Kennedy (GK) receiver is shown to closely approach the Helstrom limit. We present an experimental demonstration of this sensing task and demonstrate $15.4$ dB improvement in discrimination sensitivity over direct detection using a GK receiver, and an improvement of $19.4\%$ in error probability over coherent detection.
\end{abstract}

\section{Introduction}\label{sec:Introduction}

Optimal discrimination of quantum states has emerged as a rich and important topic in applications ranging from the measurement of qubits and quantum registers in a quantum computer \cite{wilde2013quantum} to codeword detection in photon-starved communications links \cite{chen2012optical}.  In most problems examined to date quantum state discrimination reduces to a task of discriminating between pure quantum states, relevant for the measurement of high-fidelity qubit states in a quantum register, or the discrimination of laser states in a communications channel.  However, many practical sensing tasks exist that require discrimination of mixed states, such as passive sensing or active sensing in a noisy environment.  In these scenarios traditional techniques, such as direct detection or coherent detection, are often optimal or near-optimal for discrimination when signal strengths are large, with mean photons number $\bar{n}_{S} \gg 1$ photons per mode.  When signal strengths are weak, however, these traditional sensing techniques are often sub-optimal, yielding little or no information about the received state.  In these scenarios, a full quantum mechanical treatment of the sensing problem is required to determine the fundamental quantum limit of sensitivity for measurement. 

The seminal text by Helstrom \cite{Helstrom1976} provides the framework for calculating the ultimate sensitivity with which an arbitrary set of quantum states can be discriminated from one another.  For measurement of a single copy of a quantum state, described by density matrix $\rho_{\rm i}$ drawn from a set $\{\rho_{\rm 1}$, $\rho_{\rm 2}$, ... $\rho_{\rm N} \}$, the \emph{Helstrom bound} gives the minimum average error probability for discriminating between these candidate states given a set of known prior probabilities for each individual state to occur.  The Helstrom formula gives little insight, however, into how to structure an apparatus to implement the measurement to achieve this bound.  For discrimination between pure states structured receivers have been devised \cite{Kennedy,dolinar1973,guha2011approaching,nair2014realizable,takeoka2008discrimination} and demonstrated \cite{Geremia,chen2012optical,becerra2013experimental,ferdinand2017multi,wittmann2008demonstration} in optical systems that have shown to achieve or approach the Helstrom bound.  A far less investigated topic is the discrimination of mixed states.  Yoshitani computed the Helstrom bound for the detection of weak coherent light in a bright thermal background \cite{Yoshitani} and compared the result against the classical limit.  No structured receiver was proposed, however, to beat the classical limit and approach the Helstrom bound.   More recently the quantum illumination sensing protocol \cite{tan2008quantum}, which capitalizes on the resource of quantum entanglement for state detection, has been applied for the discrimination of mixed states and a structured receiver has been proposed that can achieve the quantum Chernoff bound for discrimination of the mixed states under the two equally-likely hypotheses of the target being present or absent \cite{zhuang2017optimum}.  None of these structured receivers achieve the Helstrom bound for discriminating photon-starved mixed states at the quantum limit.

In this work we examine the simplest such problem of discrimination between a {\em coherent state}, a pure quantum state corresponding to ideal laser light, and a {\em thermal state}, a mixed quantum state, of a given optical mode.  Discrimination between coherent laser light and thermal noise light in a photon-starved regime represents a practical and important task-specific sensing challenge.  Distinguishing laser light from noise light is critical for the operation and performance of sensors such as laser warning receivers.  A laser warning receiver must distinguish a laser probe from innocuous environmental optical intensity fluctuations arising from incoherent sources, such as glint from the sun.    
\begin{figure}
\includegraphics[width = \columnwidth]{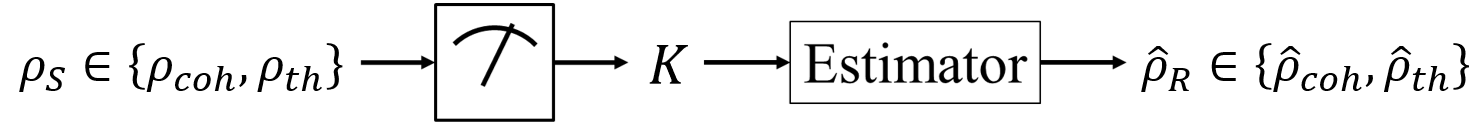}
\caption{\label{fig:receiver_schematic} Schematic of the optical receiver.  A signal $\rho_{\rm S}$ enters the receiver and is measured by a meter, resulting in a measurement result $K_{i}$.  The measurement result is passed through an estimator to determine $\hat{\rho}_{\rm R}$ which is an estimate of the most likely value of $\rho_{\rm S}$.}
\end{figure}
In the treatment of the problem considered here we represent the coherent state with density matrix $\rho_{\rm{coh}}$ and the thermal state with density matrix $\rho_{\rm th}$. For a signal with optical bandwidth $W$ Hz, a time duration of $\tau$ seconds can contain $M \sim W\cdot\tau$ orthogonal temporal modes. We pick a single spatio-temporal mode of a given polarization, whose quantum state is either $\rho_{\rm coh} = |\alpha\rangle\langle \alpha |$ (under Hypothesis 1) or $\rho_{\rm th} = \sum_{n=0}^\infty\frac{\bar{n}^{n}}{(\bar{n}+1)^{n+1}} |n\rangle \langle n|$ (under Hypothesis 2).  For the coherent state, $|\alpha\rangle = \sum_{n=0}^\infty e^{-|\alpha|^2/2}\frac{\alpha^n}{\sqrt{n!}}|n\rangle$, $|\alpha|^2 = {\bar n}$, $\alpha \in {\mathbb C}$. Here $|n\rangle$ is a Fock state (state of exactly $n$ photons) and the Fock states of $n = 0, 1, 2, \ldots, \infty$ photons span the Hilbert space of all states of a bosonic mode.  These states can be written sufficiently well in alternative bases, such as the quadrature basis, but the Fock state basis provides intuition for the experimental implementation that will be presented in section \ref{sec:experiment}.   
This sensing problem is simplified from the most general one possible in that we assume that we know (1) $\bar{n}_{\rm S}$, the mean photon number per mode in the received signal $\rho_{\rm S}$ for either hypothesis, and (2) in the event that the received state is a coherent state, we have a stable phase reference for  $\rho_{\rm coh}$.  It has been shown, recently, that relaxing such stringent assumptions still results in large improvements for quantum sensing approaches \cite{grace2019approaching}. 

\section{Quantum Bounds on Discrimination}
With a full quantum mechanical description of this binary detection problem in hand we can directly calculate the fundamental quantum limit for discrimination between the candidate signals entering the receiver.  We calculate the  error probability $P_{\rm err}$ for discriminating the single-mode states $\rho_{\rm coh}$ and $\rho_{\rm th}$, assuming equal prior probabilities. The quantum-limited minimum error probability is given by the Helstrom bound, $P_{\rm err}^{H} = \frac{1}{2}\left(1 - \frac{1}{2}||\rho_{\rm th} - \rho_{\rm coh}||_{1}\right)$ \cite{Helstrom1976}.  This fundamental limit to $P_{\rm err}$ is calculated as a function of the signal strength $\bar{n}_{S}$. The Helstrom bound provides a quantitative limit on minimum error probability but provides no guidance about the design of the optical receiver that can achieve this bound in practice.   To compare $P_{\rm err}^{\rm H}$ against bounds achievable with known receivers we consider three structured optical detection strategies and calculate their respective discrimination error probabilities for discriminating the single-mode states $\rho_{\rm coh}$ and $\rho_{\rm th}$ as a function of $\bar{n}_{S}$.  The traditional detection strategies considered are direct detection ($P_{\rm err}^{\rm DD}$) and homodyne detection ($P_{\rm err}^{\rm HD}$).  Inspired by measurement techniques that have approached optimality for coherent state discrimination problems, we also consider the Kennedy receiver ~\cite{Kennedy} ($P_{\rm err}^{\rm K}$) and the generalized Kennedy (GK) receiver, first proposed in \cite{takeoka2008discrimination}, which optimizes the coherent nulling strength ($P_{\rm err}^{\rm GK}$). For the calculations, each receiver is assumed to be operating at its quantum-noise limit.

As illustrated in fig. \ref{fig:receiver_schematic}, a receiver is comprised of a measurement strategy where a resulting measurement $K$ is provided to an estimator used to determine the most likely state $\rho_{\rm S}$ that arrived at the input to the receiver.  We consider four measurement strategies for this discrimination problem and for each strategy, we do a Bayesian analysis to generate the {\em estimator} to calculate the most likely of the two sources to have produced the receiver's measurement result.  Direct detection is assumed to be ideal mode-resolved photon number resolving detection, which reports the number of photons in the detected mode \cite{cohen2019thresholded}.  We use the photon number distributions for coherent and thermal states (an example shown in fig. \ref{fig:setup_data}(c)) to generate the estimator for discrimination.  Ideal coherent homodyne detection mixes the incoming light with a phase-referenced, strong coherent-state local oscillator (perfectly mode-matched with the signal) in a 50:50 beamsplitter, detects both outputs of the beamsplitter using ideal photodetectors (quantum-noise-limited intensity measurement), difference amplifies the two photocurrents and integrates over the mode duration.  The output of the homodyne detector is an average voltage $K = V_{\rm HD}$\cite{alexander1997optical}.  An estimator is generated by comparing $V_{\rm HD}$ against a threshold voltage $V_{\rm th}$.  If $V_{\rm HD} > V_{\rm th}$ we estimate $\hat{\rho}_{\rm S} = \hat{\rho}_{\rm coh}$.  Otherwise, we estimate $\hat{\rho}_{\rm S} = \hat{\rho}_{\rm th}$.  The value $V_{\rm th}$ is optimized to minimize the error probability for the discrimination problem.  

A diagram of the ideal implementation of the Kennedy and GK receivers is shown in fig. \ref{fig:setup_data}(a).  The GK receiver {\em displaces} the incoming signal in the phase space by $-(\alpha + \beta)$ by mixing the input light into a highly transmissive beamsplitter of transmissivity $\kappa \approx 1$ with a coherent state local oscillator $\rho_{\rm LO} = |\alpha_{\rm LO}\rangle \langle \alpha_{\rm LO}| $ with $\alpha_{\rm LO} = -(\alpha+\beta)/\sqrt{1-\kappa}$, and detects the output with an ideal photon counter; the measurment result $K$ is the number of photons measured in a single temporal mode. The displacement requires a priori knowledge of the complex amplitude $\alpha$, i.e., requires both an amplitude and phase reference of the received state $\rho_{S}$. The receiver thus displaces the coherent state $|\alpha\rangle$ to another coherent state $|-\beta\rangle$ and displaces the zero-mean thermal state to a thermal state with mean $-(\alpha+\beta)$. The Kennedy receiver uses $\beta = 0$.  For the generalized Kennedy receiver $\beta$ is chosen optimally to minimize the average error probability for discrimination.  

The average probabilities of error achieved by each of these receivers, plotted as a function of $\bar{n}_{S}$, are shown as solid lines in Fig. \ref{fig:Perr} \cite{habif2018quantum}. No structured receiver is known that can {\em exactly} attain the Helstrom limit for all values of $\bar{n}_{S}$. However, the generalized Kennedy receiver has an error probability that is within $2.4\%$ of the Helstrom bound for all values of $\bar{n}_{S}$.  A detailed explanation of the calculation of $P_{\rm err}^{\rm K}$ and $P_{\rm err}^{\rm GK}$ is given in a supplemental document to this article.

\section{Experimental Measurement}\label{sec:experiment}
\begin{figure*}
\includegraphics[width = \textwidth]{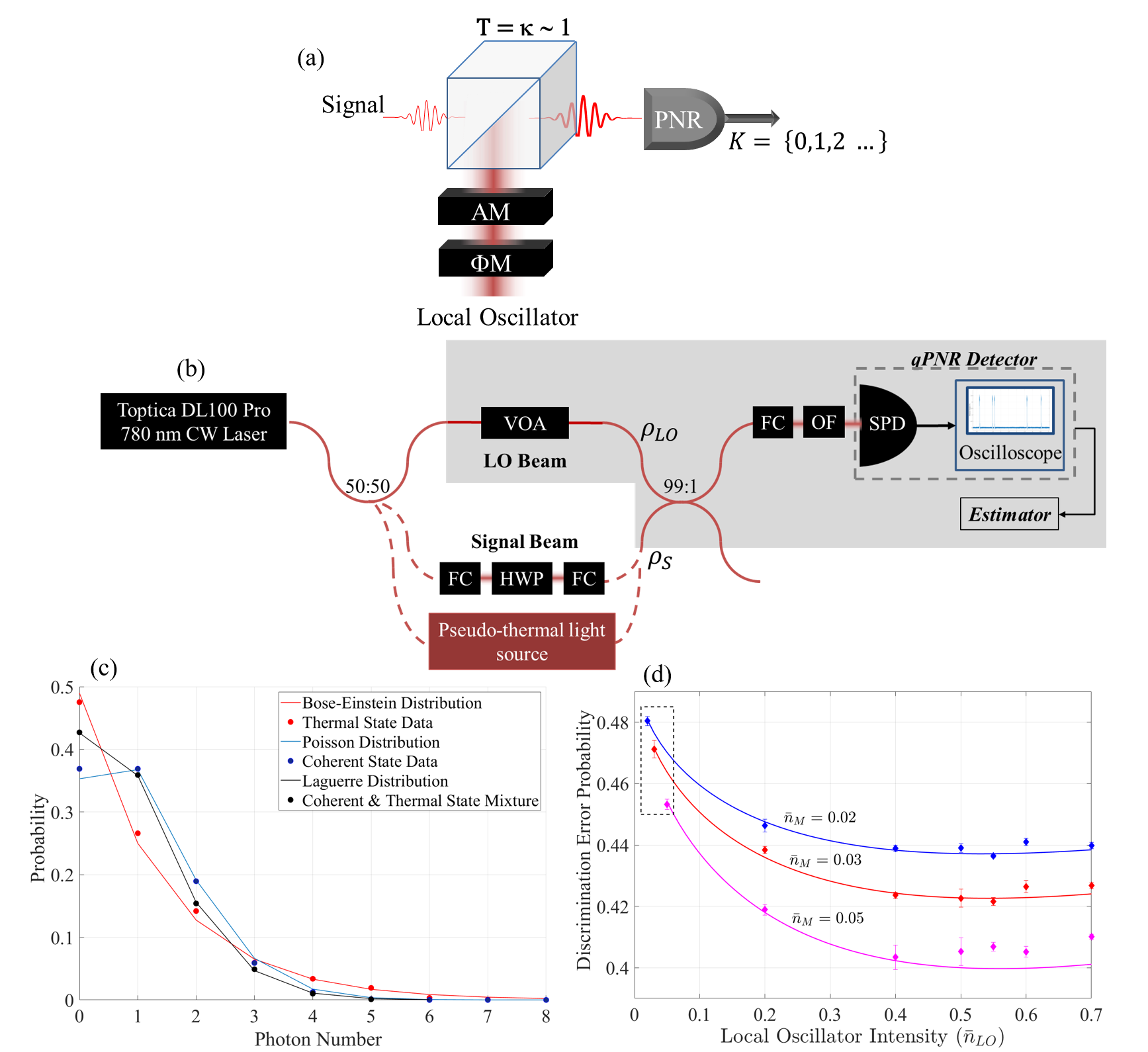}
\caption{\label{fig:setup_data} {\bf (a)} Diagram of the ideal implementation of the generalized Kennedy receiver.  The signal enters the input port of a beamsplitter with high transmissivity ($T$), where it is mixed with a local oscillator which can be optimized with phase modulation ($\Phi$M) and amplitude modulation (AM).  The resulting state is directed toward a photon number resolving detector for measurement.  {\bf (b)} Experimental setup for validating the error bounds for discrimination.  The dark gray region outlines the GK receiver.  The components identified in the design are: (VOA) variable optical attenuator, (FC) fiber collimator, (HWP) half-wave plate, (OF) optical filter and (SPD) single photon detector.  The dark red dashed lines indicate that measurements were made with $\rho_{\rm S}$ in the signal port provided by either the coherent source, or the thermal source.  We swapped between the two sources by exchanging fiber inputs to the receiver signal port.  The measured mean photon numbers of the signal beam $\bar{n}_{\rm M}$ and LO beam are calibrated by the photon detection rate of the qPNR detector. {\bf (c)} Photon counting statistics for coherent (blue) and thermal (red) light sources, and a mixture of $90\%$ coherent light with $10\%$ thermal light to generate a Laguerre distribution (black).  For all three measurements the mean photon number of the signal was $\bar{n}_{\rm M} = 1$ photon. {\bf (d)}  Discrimination error probability measured as a function of LO strength in the GK receiver, for three different signal strengths ($\bar{n}_{\rm M}$).  The data corresponding to the performance of the Kennedy receiver ($P_{\rm err}^{\rm K}$) are points on the far left of the plot corresponding to $\bar{n}_{\rm LO}=\bar{n}_{\rm M}$.  The performance of the GK receiver was measured at the data point taken at the LO intensity corresponding to theoretically predicted minimum $P_{\rm err}^{\rm GK}$.}
\end{figure*}

To validate the performance of these candidate receivers we constructed a laboratory setup to generate both coherent state and thermal state light and perform discrimination experiments implementing three of the candidate receivers.  The measurement setup for experimentally validating our error discrimination bounds is shown in figure \ref{fig:setup_data}(b).  A narrow linewidth, continuous wave laser (Toptica DL100 Pro), operating at $\lambda = 780$ nm is used as the optical source for the experiment.  To generate both the signal $\rho_{\rm S}$ and the local oscillator reference  $\rho_{LO}$ for the Kennedy and GK receiver, the pump laser is coupled to a $50:50$ fiber beamsplitter generating two output beams. To control the amplitude of the local oscillator a variable fiber optical attenuator (VOA) is used allowing control of the amplitude $\beta$ to minimize the final detection error probability.

To prepare a coherent state signal ($\rho_{\rm S} = \rho_{\rm coh}$) the signal beam is sent through a half-wave plate in free space where we set the polarization to match that of the LO. 
To generate the thermal state ($\rho_{\rm S} = \rho_{th}$) we constructed a setup following \cite{koczyk1996photon,stangner2017step} to create a narrowband source of pseudo-thermal light. The pump laser illuminated a rotating diffuse reflector and the scattered light from the diffuser was collected by a lens coupling it to a single-mode optical fiber directed toward the signal arm of our experimental setup.  The overall efficiency of our pseudothermal source was very low ($~-100$ dB), due to the weak coupling of the diffuse light into the optical fiber.  The polarization of the light from the laser source was preserved through the scattering process, and a half wave plate was used to adjust the polarization direction of the pseudothermal state.  The selective spatial filtering from the single-mode optical fiber resulted in residual coherence in the thermal state which was accounted for during the local oscillator mixing process during state preparation.  

The Kennedy and GK receivers are implemented by mixing the signal and LO beams within a highly asymmetric beamsplitter to faithfully emulate the design in fig. \ref{fig:setup_data}(a).  We use a $99:1$ fiber beamsplitter, with $\rho_{\rm S}$ injected into the highly transmissive port of the beamsplitter and $\rho_{\rm LO}$ injected into the highly reflective port.  The resulting state is directed toward an optical bandpass filter, with bandwidth $\delta \lambda = 10$ nm to reject environmental noise before irradiating a single photon detector (SPD).  We characterized the mode matching that could be achieved within the Kennedy receiver and measured an extinction $>18$ dB when the intensities of $\rho_{\rm LO}$ and $\rho_{\rm S}$ were matched.  This imperfect extinction is likely due to polarization mismatch between the modes and ultimately limits the performance of the GK receiver for discrimination.  

A quasi-photon number resolving (qPNR) detector was implemented by making time-resolved measurements of single photon events within a $\tau_{\rm s} = 1\mu s$ measurement window.  A Perkin-Elmer Si avalanche photodiode SPD (SPCM-AQR-14) was used to count photon detection events.  The minimum detection time (1/maximum counting rate) for this detector is $\tau_{\rm D}\sim 50$ ns.  By ensuring that $\zeta = \frac{\tau_{D}}{\tau{s}} << 1$ and $\zeta\cdot\bar{n}_{\rm s} << 1$, a detection event by the SPD can be approximated as a projection onto the $|n=1\rangle$ Fock basis state within the time interval $\tau_{\rm D}$.  The output from the SPD was directed toward an oscilloscope (Tektronix MDO4104) which records the number of single photon detection events that are made within $\tau_{\rm s}$ and yields a measurement projecting onto the Fock basis $|n = \hat{n}_{i}\rangle$, which is used as the measurement result $K_{i}$ in our estimator (fig. \ref{fig:receiver_schematic}).  The background counting rate at the SPD, when no signal was present at the detector, was measured to be $4\cdot10^{-4}$ counts per microsecond, which has negligible effect in our calculation of discrimination error probability.  The overall receiver system efficiency was measured to be $\eta_{\rm D} = 0.45$.  This efficiency accounts for losses in fiber components and connectors, but is dominated by the inefficiency of the SPD. In order to compare our measurements directly against the Helstrom bound for this discrimination problem, we define the mean photon number for both the LO and signal by the detected photon rate within the qPNR detector.  The quantity $\bar{n}_{\rm M} = \eta_{\rm D}\cdot\bar{n}_{\rm S}$ is the measured mean photon rate of the signal, and $ \bar{n}_{\rm LO}$ is the measured mean photon rate of the LO beam at the qPNR detector.  The Kennedy receiver requires that these two intensities are well-matched at the qPNR detector and the GK receiver requires precise control of their ratio. 
    
\subsection{Experimental Results}
The plots shown in Fig. \ref{fig:setup_data}(c) and \ref{fig:setup_data}(d) illustrate representative data for the direct detection receiver and the Kennedy and GK receivers, respectively.  To characterize the direct detection receiver, $\rho_{\rm coh}$ and $\rho_{\rm th}$ were each directed toward the qPNR detector, and photon counting statistics were collected over $1000$ measurement trials.  Figure \ref{fig:setup_data}(c) shows the probability distributions for photon detection measurements for coherent state light (blue dots) and thermal state light (red dots) with corresponding theoretical distributions of Poisson and Bose-Einstein statistics, respectively.  The data shows excellent agreement with the theoretical predictions\cite{bondurant1982photon}.  Next, we implemented the Kennedy and GK receivers by mixing the signals with the coherent state LO prior to detection with the qPNR detector.  Photon detection events are recorded from the qPNR detector and used to make an estimate of the state received.  When the state $\rho_{\rm coh}$ is mixed with $\rho_{\rm LO}$ in the receiver beamsplitter, the resulting output state is also a coherent state, exhibiting Poisson distributed photostatistics.  When the thermal state $\rho_{\rm th}$ is mixed with $\rho_{\rm LO}$ the resulting photostatistics exhibit a Laguerre distribution \cite{Helstrom1976,alexander1997optical}.  Figure \ref{fig:setup_data}(c) shows a measured Laguerre distribution (black dots) and theoretical fit (black line) of the qPNR statistics on a single-mode displaced thermal state of mean photon number $\bar{n}_{\rm M} = 1$, obtained by applying a real-valued coherent displacement $\beta$, $|\beta|^2 = 0.9$ to a zero-mean thermal state with mean photon number $\bar{n}_{\rm th} = 0.1$. The error probability $P_{\rm err}$ is computed from these normalized experimental distributions.  For a value $\bar{n}_{\rm S}$ we calculate the two photon counting distributions $P\left(\hat{\rho}_{\rm R} = \rho_{\rm S}|K = n\right)$, for each receiver and we use these distributions to implement our estimator.  For a single measurement $K = n$ we determine the most likely distribution from which this measurement was drawn and estimate the received state as $\hat{\rho}_{\rm R} = \rho_{\rm S}$.  The discrimination error probability is calculated as the fraction of times this decision results in an error.  

\begin{figure*}
\includegraphics[width = \textwidth]{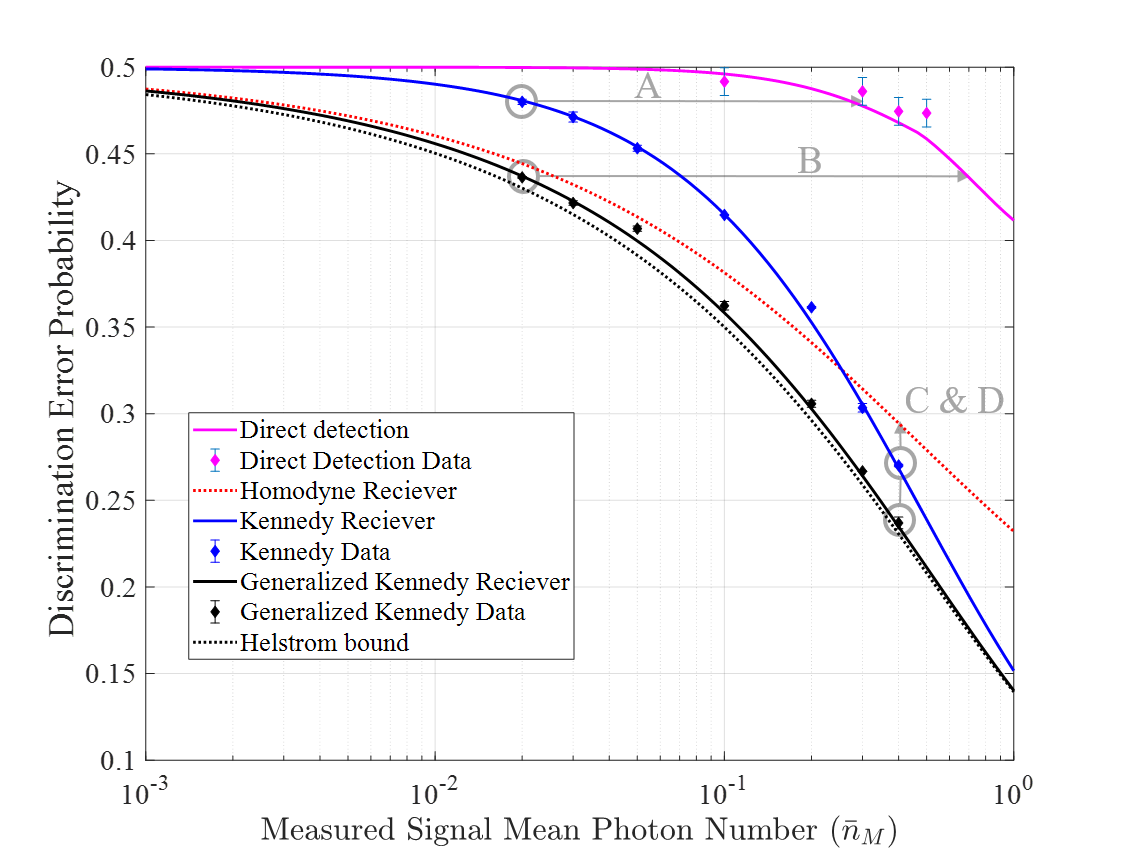}
\caption{\label{fig:Perr} Theoretic discrimination error probability curves calculated in \cite{habif2018quantum}, compared with experimental data for the direct detection receiver with the qPNR detector, the Kennedy receiver with the SPD and the generalized Kennedy receiver using the qPNR detector.  The error rates are plotted as a function of the measured signal mean photon number $\bar{n}_{\rm M}$. The error bars are computed from calculating the standard deviation from 5 independent trials. The error bars for the Kennedy and GK receivers error rates are smaller than the data marker used for plotting.}
\end{figure*}

The results of our experimental measurements for three different receiver implementations are shown in figure \ref{fig:Perr}.  The direct detection receiver used the qPNR detector to directly count photons from the states $\rho_{\rm coh}$ and $\rho_{\rm th}$ as a function of the mean photon numbers of the states.  Collecting $N = 10^{3}$ one microsecond trials detecting $\rho_{\rm coh}$ and $\rho_{\rm th}$ we constructed experimental photon counting distributions, such as those shown in figure \ref{fig:setup_data}(c).  The theoretical and measured error probability for direct detection is shown in the magenta solid line and diamonds, respectively, in fig. \ref{fig:Perr}.
For the Kennedy and GK receivers, the strategy for generating the estimator is unchanged, but the photon counting statistics must account for the mixing of the LO beam with the signal beam in the beamsplitter.  When $\rho_{\rm S} = \rho_{\rm coh}$ mixes with the LO, the output is a displaced coherent state with Poisson statistics.  When $\rho_{\rm S} = \rho_{\rm th}$ mixing with the coherent state LO generates photon statistics with a Laguerre distribution\cite{Helstrom1976,alexander1997optical}.  The mixing is followed by direct detection by the qPNR detector.  The optimal estimator for both the Kennedy and GK receivers decides $\rho_{\rm S} = \rho_{\rm coh}$ when $K = 0$, and $\rho_{\rm S} = \rho_{\rm th}$ when $K > 0$.  This estimator obviates the need for photon number resolving capability in our receiver, as the estimator can be implemented using a simple threshold detecting SPD.  Interestingly, photon number resolution has been shown as an important capability in multi-symbol ($>2$) coherent state discrimination \cite{izumi2013quantum} as well as mixed state sensing \cite{cohen2019thresholded}, but for our binary discrimination problem it is not necessary.   

Figure \ref{fig:setup_data}(d) shows the theoretical and measured error probabilities, as a function of the LO intensity, for three different values of signal strength: $\bar{n}_{\rm M} = 0.02, 0.03$ and $0.05$ photons per mode. In fig. \ref{fig:Perr}, the data point corresponding to $\bar{n}_{\rm LO} = \bar{n}_{\rm M}$ is the measured error rate for the Kennedy receiver $P_{\rm err}^{\rm K}$.  As $\bar{n}_{\rm LO}$ is increased the error probability decreases until it reaches a minimum error rate.  This minimum error $P_{\rm err}^{\rm GK}$ is the optimal operating point for the GK receiver.  We selected the measured minimum error as the data point corresponding to the theoretically predicted value of $\bar{n}_{\rm LO}$ that yields the minimum error.  As $\bar{n}_{\rm LO}$ is increased beyond the value achieving $P_{\rm err}^{\rm GK}$, the measured error probability increased more steeply than predicted by the theory (solid curves).  This is likely due to imperfect phase matching between the LO and the signal, seen previously in \cite{chen2012optical}.  

The blue diamonds in figure \ref{fig:Perr} are the experimental results $P_{\rm err}^{\rm K}$ for the Kennedy receiver, closely matching the theoretical prediction (solid blue line).  The black diamonds are the experimental results $P_{\rm err}^{\rm GK}$ plotted against the theoretical prediction for the performance of the GK receiver (black solid line).  The measured $P_{\rm err}^{\rm GK}$ closely matches the theoretical predicted performance of the receiver.  

At all points over our measured range of $\bar{n}_{\rm M}$, $P_{\rm err}^{\rm GK} < P_{\rm err}^{\rm K} < P_{\rm err}^{\rm DD}$.  We show that the Kennedy and GK receivers are measured to outperform the calculated sensitivity of a homodyne receiver at $\bar{n}_{M} = 0.25$ and $\bar{n}_{M} = 0.02$, respectively.  At a value $\bar{n}_{M} = 0.02$ photons, the Kennedy and GK receivers deliver $11.7 dB$ and $15.4 dB$ improvement, respectively in receiver sensitivity over that achieved by direct detection (labelled as $A$ and $B$, respectively, in fig. \ref{fig:Perr}).  At a value $\bar{n}_{M} = 0.4$ photons, the Kennedy and GK receivers are demonstrated to improve on $P_{\rm err}^{\rm HD}$ by $8.1\%$ and $19.4\%$, respectively (labelled in fig. \ref{fig:Perr} as $C$ and $D$).  



\begin{figure*}
    \centering
    \includegraphics[width = 4in]{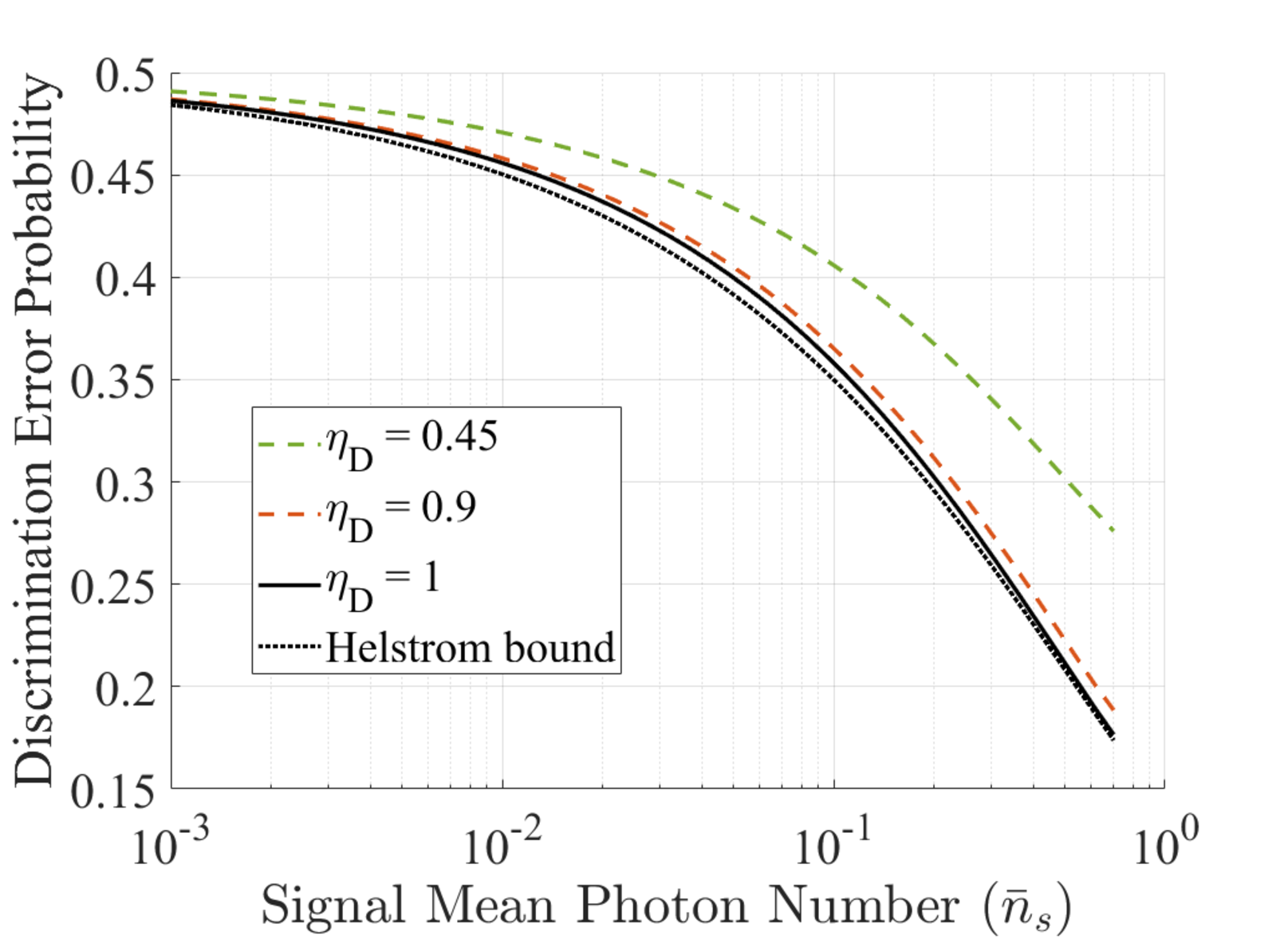}
    \caption{Performance of the GK receiver for three different values of receiver efficiency (insertion loss).}
    \label{fig:efficiency}
\end{figure*}
Finally, we discuss the impact of imperfect receiver efficiency on the performance of the optical receiver.  Our experiment utilized a cost-effective, commercially available SPD to realize a receiver with efficiency $\eta_{\rm D} = 0.45$.  However, commercially available superconducting single photon detection technology can achieve $\eta_{\rm D} > 0.90$.  Detection efficiencies $\eta_{\rm D} > 0.95$ have been demonstrated for superconducting nanowire \cite{Lita:08} and transition edge sensor \cite{tsujino2011quantum} detector technologies. To provide an understanding of the impact of receiver efficiency on discrimination error probability, we show $P_{\rm err}$ curves for the GK receiver with $\eta_{\rm D} = 0.45, 0.90, \, \text{and}\, 1.0$ in fig. \ref{fig:efficiency}.

\section{Discussion} \label{sec:discussion}
In this work, we considered the problem of discriminating the coherent state $|\alpha\rangle$ and the thermal state $\rho_{\rm{th}}$ of a given optical mode, of a given mean photon number ${\bar n}_{\rm S}$ per mode. We evaluated the Helstrom (quantum) limit of minimum probability of error achievable with any physically-permissible receiver design. Further, we proposed the use of the generalized Kennedy (GK) receiver for this discrimination problem, where previously it had been reserved for discrimination problems involving only pure coherent states. The GK receiver applies a coherent amplitude displacement of $-(\alpha + \beta)$ to the quantum state of the received mode prior to detecting it using photon number resolving (PNR) detection. Instead of `nulling' the coherent state hypothesis (displacing it to vacuum) as the original Kennedy receiver did, the GK receiver over-displaces it to the coherent state $|-\beta\rangle$, if the coherent state $|\alpha\rangle$ was received. If the received mode was excited in the (zero-mean) thermal state, the GK receiver displaces it to a thermal state of mean $-(\alpha + \beta)$. If one were to detect the optical mode directly (without applying the displacement) using PNR detection, the coherent state would produce a random number of clicks that follows a Poisson distribution of mean ${\bar n}_{\rm S} = |\alpha|^2$, whereas the thermal state would produce a number of clicks following a Bose-Einstein (geometric) distribution of mean ${\bar n}_{\rm S}$. For the GK receiver, which does PNR detection {\em after} the displacement is applied, the coherent state hypothesis produces clicks with a Poisson distribution of mean $|\beta|^2$, whereas the thermal state hypothesis produces a number of clicks that follows a Laguerre distribution. We showed that the GK receiver outperforms ideal direct (PNR) detection, and approaches the Helstrom limit very closely. The GK receiver is asymptotically optimal as ${\bar n}_{\rm S} \to \infty$, yet, its error probability is within $2.4\%$ of the Helstrom limit for all ${\bar n}_{\rm S}$.  

We realized the GK receiver in a proof-of-concept experiment. We characterized the pre-displacement and post-displacement PNR detection statistics for both hypotheses, and corroborated with theory. We then demonstrated the improvement in the average error probability, by using the ${\bar n}_{\rm S}$-dependent optimized displacement amplitude. The experimental data in fig.\ref{fig:setup_data}(d) shows that the error probability improvement provided by the GK receiver is quite robust against fluctuations in the LO amplitude compared to the Kennedy receiver, as the value $\tfrac{d}{d \rm {\bar n}_{\rm LO}}\left(P_{\rm err}^{\rm GK}\right) \longrightarrow 0$ at the optimum operating point.  Phase mis-match for the Kennedy and GK receivers are analyzed in \cite{chen2012optical} for pure state discrimination.  We expect that the sensitivity of our discrimination task to phase fluctuations will be reduced, since the measurement result from a thermal state ($\rho_{\rm th}$) is unchanged under fluctuations of the LO beam. 

One important follow on for this work is to extend this receiver concept for the general problem of {\em classifying light}, one that has numerous applications ranging from astronomical imaging, fluorescence microscopy and remote sensing, to tactical applications. The first step in doing so is to relax the assumptions on the prior information about the quantum state of the optical signal. One possibility to explore is an adaptive receiver where the first stage of the receiver uses a conventional optical receiver (e.g., PNR or homodyne detection) to do partial tomography of the unknown signal, and adapts to a second stage that applies a pre-detection optical transformation (e.g., displacement, and squeezing) that is optimized to the partial knowledge about the signal acquired during the first stage. 

While the GK receiver approaches, but does not exactly meet, the Helstrom bound for error probability for the single-shot state discrimination problem we consider here, we showed that it is optimal for achieving the quantum Chernoff bound exactly~\cite{habif2018quantum}, a quantity that characterizes the optimal error probability exponent for multi-copy state discrimination. An interesting topic for investigation is why different receiver designs are optimal for single-copy and multi-copy state discrimination. Examination of this sensing problem is an initial foray into advances in general sensing and discrimination of multi-mode mixed bosonic states, with practical applications to  a wide range of problems in photon-starved photonic sensing scenarios.

\section*{Funding}

Work at USC  was sponsored by Army Research Office Grant Number W911NF-19-1-0351.  Work at U of A was sponsored by an Office of Naval Research 6.2 project on photonic quantum sensing technologies, awarded under contract number N00014-19-1-2189.

\section*{Acknowledgments}
The views and conclusions contained in this document are those of the authors and should not be interpreted as representing the official policies, either expressed or implied, of the Army Research Office or the U.S. Government. The U.S. Government is authorized to reproduce and distribute reprints for Government purposes notwithstanding any copyright notation herein.

The authors thank William Oliver and Franco Wong, from MIT, for the generous loan of equipment for the experiment.

\section*{Disclosures}
The authors declare no conflicts of interest.

\bibliography{JLH_OE_refs}






\end{document}


\maketitle

\section{Problem setup}

\begin{figure}
    \centering
    \includegraphics[width = 5in]{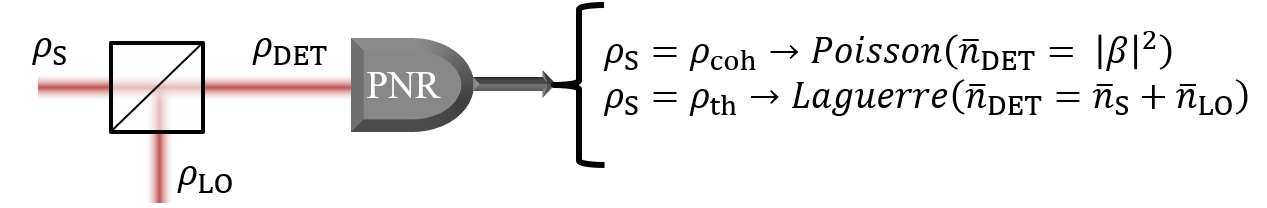}
    \caption{Schematic of the Kennedy and generalized Kennedy receiver architecture.}
    \label{fig:architecture}
\end{figure}

A schematic of the Kennedy and generalized Kennedy receiver architecture is shown in fig. \ref{fig:architecture}.  The signal state entering the receiver is $\rho_{\rm S} \in (\rho_{\rm coh},\rho_{\rm th})$, where the quantum states for $\rho_{\rm coh}$ and $\rho_{\rm th}$ are presented in the main paper in the Fock basis.  The received state $\rho_{\rm S}$ is mixed with a local oscillator beam within a highly transmissive beamsplitter with transmissivity $\kappa \sim 1$.   The state $\rho_{\rm S}$ has a mean photon number $\bar{n}_{\rm S} = |\alpha|^{2}$.  The local oscillator state $\rho_{\rm LO}$ is a pure coherent state written $\rho_{\rm LO} = |\alpha_{\rm LO}\rangle \langle \alpha_{\rm LO}| $ with $\alpha_{\rm LO} = -(\alpha+\beta)/\sqrt{1-\kappa}$, and mean photon number $\bar{n}_{LO} = |\alpha+\beta|^{2}/(1 - \kappa)$.  For the Kennedy receiver $\beta = 0$.  For the GK receiver $\beta$ is optimized to minimize $P_{\rm err}^{\rm GK}$.  After mixing on the beamsplitter, the output state ($\rho_{\rm DET}$) is sent to an ideal photon number resolving (PNR) detector for measurement with direct detection, the output of which reports the number of photons ($n_{DET} = 0,1,2 ...$) detected in $\rho_{\rm DET}$.

\section{Discrimination error probability}

If $\rho_{\rm S} = \rho_{\rm coh}$ the state at the output of the beamsplitter is a displaced coherent state $\rho_{\rm DET} \longrightarrow |\beta\rangle\langle\beta|$, and direct detection on this state results in Poisson distributed counting statistics with mean photon number $\bar{n}_{\rm DET} = |\beta|^{2}$.  If $\rho_{\rm S} = \rho_{\rm th}$ the state at the output of the beamsplitter is a displaced thermal state with mean photon number $\bar{n}_{\rm DET} =\bar{n}_{LO} + \bar{n}_{S}$.  Direct detection of this state results in counting statistics that follow a Laguerre distribution described in \cite{alexander1997optical}.

The error probability for discrimination of thermal and coherent states, with equal prior probabilities is written
\begin{equation}
\label{eq:errprob}
    P_{err} = \tfrac{1}{2}\left( P\left(\hat{\rho}_{\rm coh}|\rho_{\rm S} = \rho_{\rm th}\right) + P\left(\hat{\rho}_{\rm th}|\rho_{\rm S} = \rho_{\rm coh}\right)\right),
\end{equation}
where a hat over a density operator denotes our estimate of the received state.  In eq. \ref{eq:errprob} $P\left(\hat{\rho}_{\rm coh}|\rho_{\rm S} = \rho_{\rm th}\right)$ is the probability of estimating that $\hat{\rho}_{\rm coh}$ was received when $\rho_{\rm S} = \rho_{\rm th}$, and $P\left(\hat{\rho}_{\rm th}|\rho_{\rm S} = \rho_{\rm coh}\right)$ is the probability of estimating that $\hat{\rho}_{\rm th}$ was received when $\rho_{\rm S} = \rho_{\rm coh}$. 

The Bayesian estimators that we use in conjunction with the Kennedy and GK receivers to implement this discrimination are drawn from the counting statistics from the Poisson and Laguerre distributions described above, and in the main paper.  For both the Kennedy and GK receiver configurations we estimate $\hat{\rho}_{\rm coh}$ if $n_{DET} = 0$, and we estimate $\hat{\rho}_{\rm th}$ otherwise.  Discrimination errors are made when we detect $n_{DET} = 0$ when $\rho_{\rm S} = \rho_{\rm th}$, or we detect $n_{DET} > 0$ when $\rho_{\rm S} = \rho_{\rm coh}$.  The conditional error probabilities in eq. \ref{eq:errprob} are then calculated as
\begin{align}
    &P\left(\hat{\rho}_{\rm coh}|\rho_{\rm S} = \rho_{\rm th}\right) = P\left(n_{DET} = 0|\rho_{\rm S} = \rho_{\rm th}\right) \\
    &P\left(\hat{\rho}_{\rm th}|\rho_{\rm S} = \rho_{\rm coh}\right) = P\left(n_{DET} > 0|\rho_{\rm S} = \rho_{\rm coh}\right),
\end{align}
where $P\left(n_{DET}|\rho_{\rm S} = \rho_{\rm coh}\right)$ and $P\left(n_{DET}|\rho_{\rm S} = \rho_{\rm th}\right)$ are given by the Poisson and Laguerre distributions, respectively.  The error probability for the Kennedy receiver ($P_{\rm err}^{\rm K}$) is calculated when $\bar{n}_{LO} = \bar{n}_{S}$.  To compute the error probability for the GK receiver ($P_{\rm err}^{\rm GK}$) for each value of signal intensity $\bar{n}_{S}$ we sweep the intensity of the LO beam $\bar{n}_{LO}$ to find the minimum of eq. \ref{eq:errprob}.


\bibliography{JLH_OE_refs}
